\documentclass{ws-p8-50x6-00}
\begin{document}

\title{Hadronic Shower Development in Tile Iron-Scintillator Calorimetry}

\author{Y.A.\ Kulchitsky}

\address{for TILECAL Collaboration \\[1mm]
JINR, Dubna, Russia \& IP National Academy of Science, Minsk, Belarus
\\E-mail: Iouri.Koultchitski@cern.ch}

\maketitle

\abstracts{
The lateral and  longitudinal profiles of hadronic showers detected by
a prototype of the ATLAS Iron-Scintillator Tile Hadron Calorimeter
have been investigated.
This calorimeter uses a unique longitudinal configuration of 
scintillator tiles.
Using a fine-grained pion beam scan at 100 GeV, a detailed picture of  
transverse shower behavior is obtained.
The underlying radial energy densities for four depth segments and for 
the entire calorimeter have been reconstructed.
A three-dimensional hadronic shower parametrization has been developed.
The results presented here are useful for understanding the performance 
of iron-scintillator calorimeters, 
for developing fast simulations of hadronic showers, 
for many calorimetry problems requiring the integration of a shower 
energy deposition in a volume and for future calorimeter design.}  

\section{The Calorimeter}

We report on an experimental study of hadronic shower profiles detected by
the prototype of the ATLAS Barrel Tile Hadron Calorimeter 
(Tile calorimeter) 
\cite{tilecal-tdr96}.
The innovative design of this calorimeter, using longitudinal segmentation 
of active and passive layers 
provides an interesting system for the 
measurement of hadronic shower profiles.
Specifically, we have studied the transverse development of hadronic 
showers 
using 100 GeV pion beams and longitudinal development of hadronic showers 
using 20 -- 300 GeV pion beams.
\begin{figure*}[tbph]
     \begin{center}
        \begin{tabular}{|c|}
        \hline
\mbox{\epsfig{figure=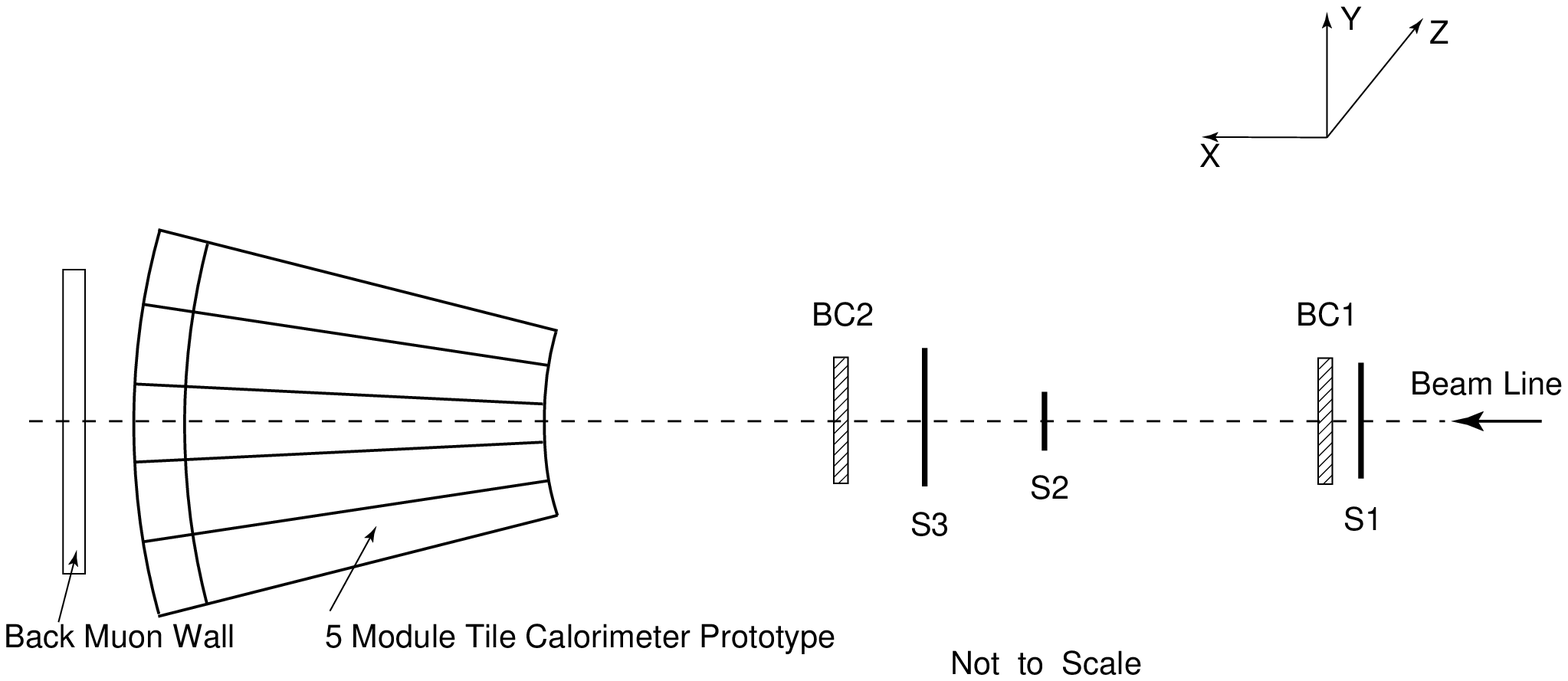,width=0.6\textwidth,height=0.2\textheight}}
\\
        \hline
        \end{tabular}
     \end{center}
       \caption{
       Schematic layout of the experimental setup (side view).
       \label{fig:f01}}
\end{figure*}
The prototype Tile Calorimeter used for this study is composed 
of five modules stacked in the $Y$ direction, 
as shown in Fig.\ \ref{fig:f01}.
Each module spans $2 \pi / 64$ in the 
azimuthal angle, 100 cm in the $Z$ direction,
180 cm in the $X$ direction (about 9 interaction lengths, 
$\lambda_I$, or about 80 
effective radiation lengths, $X_{0}$), and has a front face of 100 
$\times$ 20 cm$^2$
\cite{berger95}.
The iron to scintillator ratio is $4.67 : 1$ by volume.
The modules are divided into five segments along $Z$ and 
they are also longitudinally segmented (along $X$) into four depth segments.
The readout cells have a lateral dimensions of 200 mm along $Z$, 
and longitudinal 
dimensions of 300, 400, 500, 600 mm for depth segments 1 -- 4.
The calorimeter was placed on a scanning table that allowed
movement in any direction.
Upstream of the calorimeter, a trigger counter telescope (S1, S2, S3) 
was installed, defining a beam spot approximately 20 mm in diameter.
Two delay-line wire chambers (BC1 and BC2), each with $(Z, Y)$  readout,
allowed the impact point of beam particles on the calorimeter face to be 
reconstructed to better than $\pm 1$ mm
\cite{ariztizabal94}.
``Muon walls'' were placed behind (800$\times$800 mm$^2$), 
shown in Fig.\ \ref{fig:f01} as ``Back Muon Wall'', 
and on the positive $Z$ side (400$\times$1150 mm$^2$),
not seen in Fig.\ \ref{fig:f01}, of the calorimeter
modules to measure longitudinal and lateral hadronic shower leakage.
The data used for the study of lateral profiles were collected 
in 1995 during a special $Z$-scan run at the CERN SPS test beam.  
The calorimeter was exposed to 100 GeV negative pions at a $10^{\circ}$ 
angle with varying impact points in the $Z$-range from $- 360$ to $200$ mm.
A total of $>$~300,000 events have been analysed.
The uniformity of the calorimeter's response for this $Z$-scan is
estimated to be $1\%$  
\cite{budagov96-76}.
The data used for the study of longitudinal profiles were obtained using 
20 -- 300 GeV negative pions at a $20^{\circ}$ angle and were also taken
in 1995 during the same test beam run. 

\section{Extracting the Underlying Radial Energy Density}

There are several methods for extracting the radial density $\Phi(r)$
from the measured distributions of energy depositions.
One method was used in the analysis of the data from the lead-scintillating 
fiber calorimeter \cite{acosta92}.
Another method for extracting the radial density 
is to use the marginal density function $f(z)$
which is related to the radial density $\Phi(r)$ \cite{budagov97},
\begin{equation}
        \Phi(r) = 
                - \frac{1}{\pi}\ \frac{d}{dr^2}
                \int_{r^2}^{\infty} 
                \frac{f(z)\ d z^2}{\sqrt{z^{2}-r^{2}}}.
\label{e5}
\end{equation}
We used the sum of three exponential functions to
parameterize $f(z)$ as
\begin{equation}
        f(z)  = 
                \frac{E_{0}}{2B}\  \sum_{i=1}^{3}\  a_i\ 
                e^{ - \frac{|z|}{{\lambda}_{i}} }, 
\label{e21}
\end{equation}
where 
$z$ is the transverse coordinate,
$E_{0},\ a_i,\ \lambda_i$ are free parameters, 
$B =  \sum_{i=1}^{3} a_i \lambda_i$, 
$\sum_{i=1}^{3} a_i = 1$
and $\int_{-\infty}^{+\infty} f  (z) dz = E_{0}$.
The radial density function, obtained by integration 
and differentiation of equation (\ref{e5}), is 
\begin{equation}
        \Phi(r) = 
                \frac{E_{0}}{2 \pi B}\ 
                \sum_{i=1}^{3}\ \frac{a_{i}}{\lambda_{i}}\
                K_{0} (r/\lambda_{i}), 
\label{e23}
\end{equation}
where 
$K_{0}$ is the modified Bessel function.
We define a column of five cells in a depth segment as a tower. 
Using the parametrization shown in equation (\ref{e21}),
we can show that the energy deposition in a tower
can be written  as
\begin{eqnarray}
        E(z)  = 
&               E_{0} - 
               \frac{E_{0}}{B}\  \sum_{i=1}^{3}\  a_i {\lambda}_i\
                \cosh (\frac{ |z| }{ {\lambda}_i }) \
                e^{ - \frac{h}{2 {\lambda}_i} },  
&  for\         |z| \leq \frac{h}{2},
\\
        E(z) =  
&              \frac{E_{0}}{B}\ \sum_{i = 1}^{3}\ a_i {\lambda}_i\
                \sinh (\frac{h}{2 \lambda_i }) \ 
                e^{ - \frac{|z|}{{\lambda}_i}},
&  for\         |z| > \frac{h}{2},
\label{e023}
\end{eqnarray}
where $h$ is the size of the front face of the tower along the $z$ axis.

\section{Transverse Behaviour of Hadronic Showers}

\begin{figure*}[tbph]
     \begin{center}
        \begin{tabular}{cc}
\mbox{\epsfig{figure=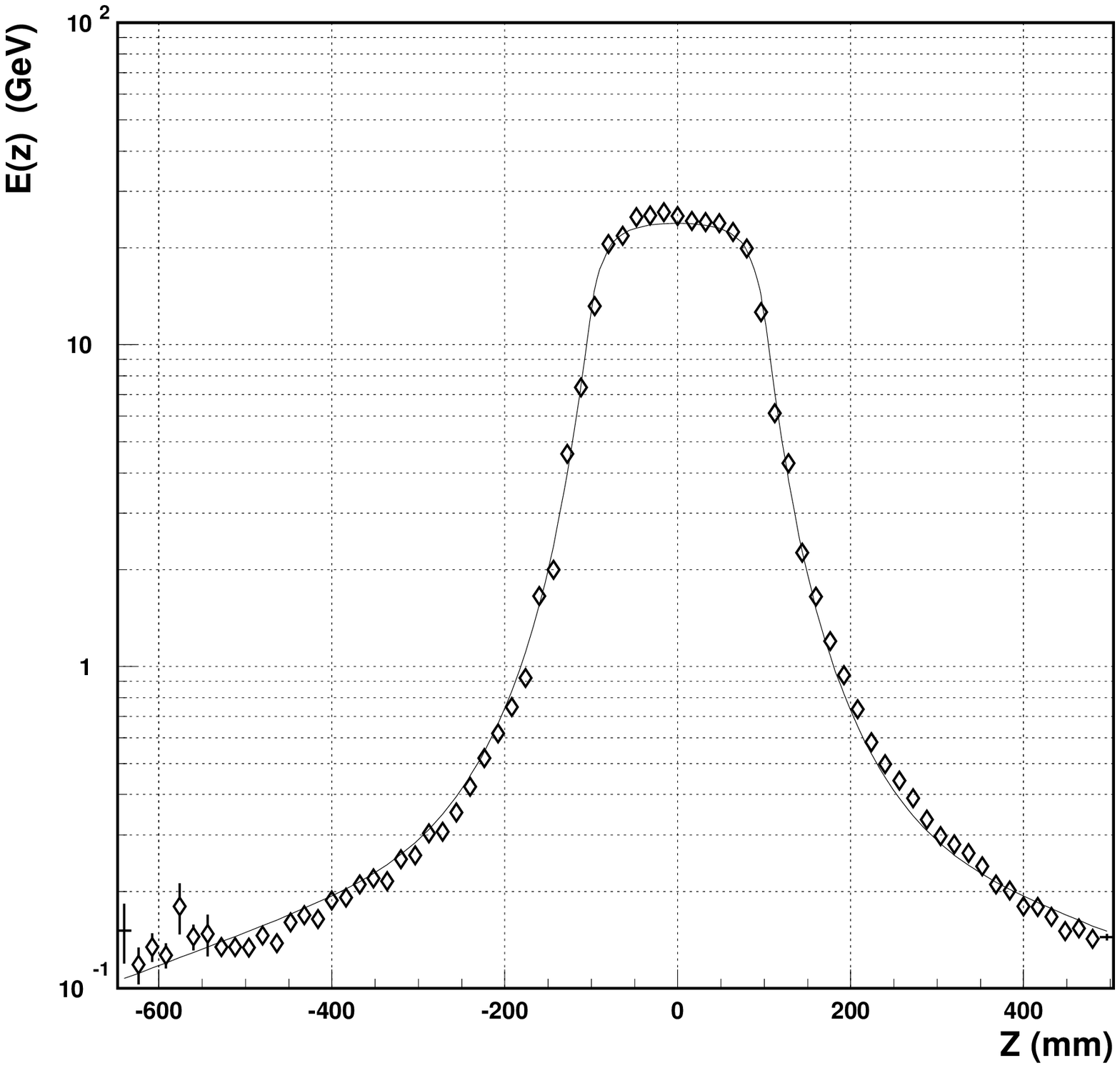,width=0.4\textwidth,height=0.2\textheight}}
&
\mbox{\epsfig{figure=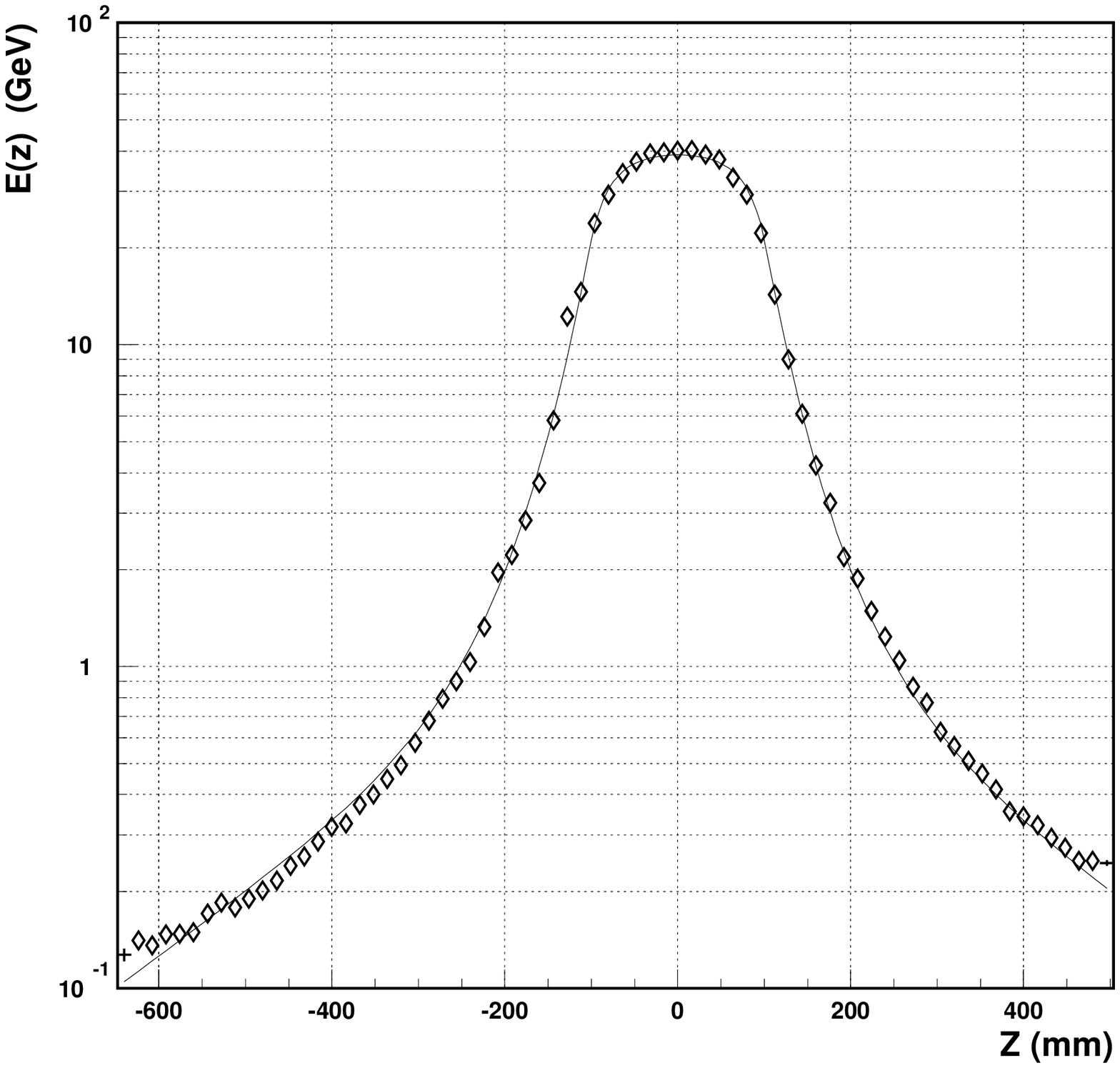,width=0.4\textwidth,height=0.2\textheight}}
        \\
\mbox{\epsfig{figure=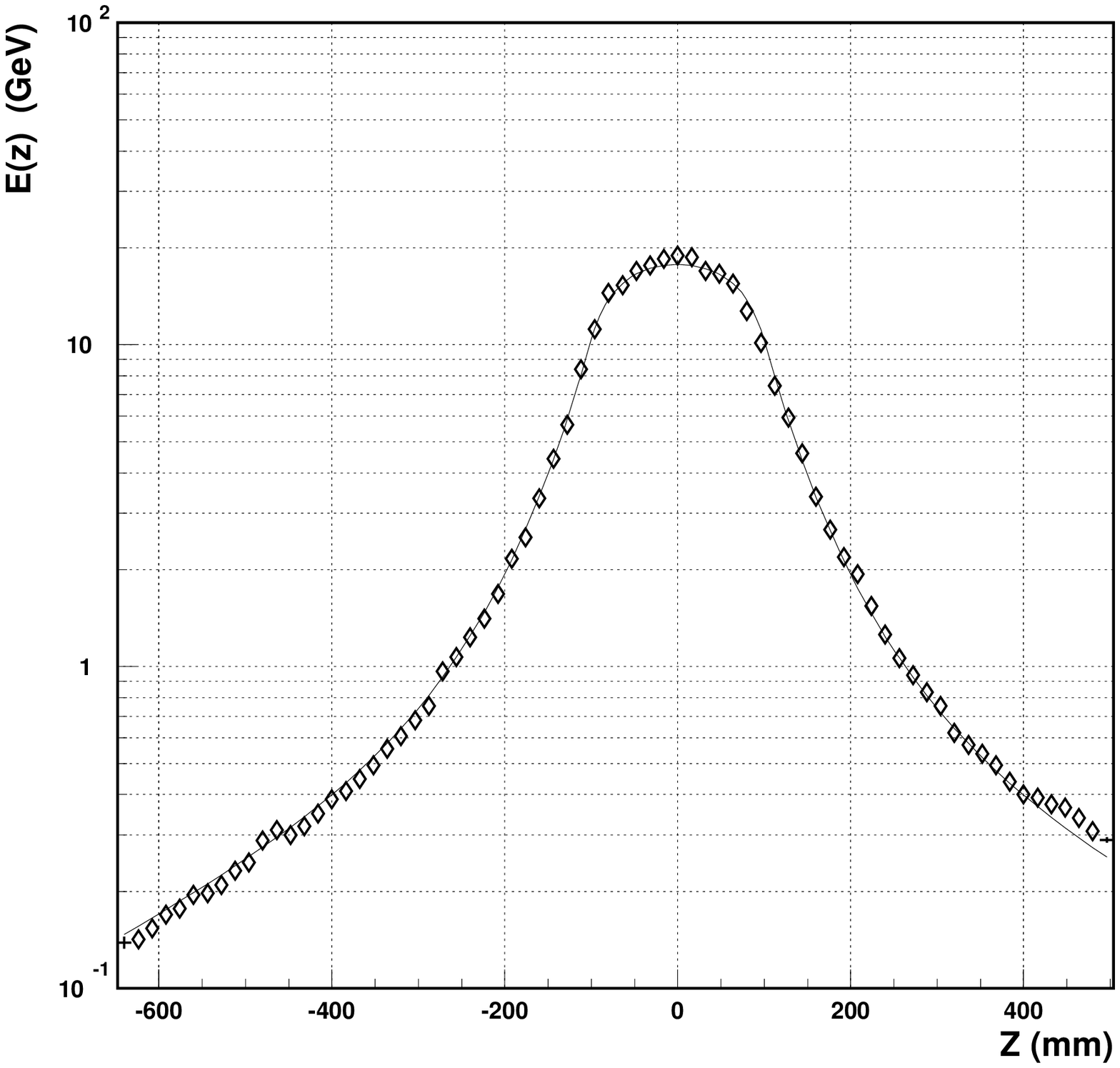,width=0.4\textwidth,height=0.2\textheight}}
&
\mbox{\epsfig{figure=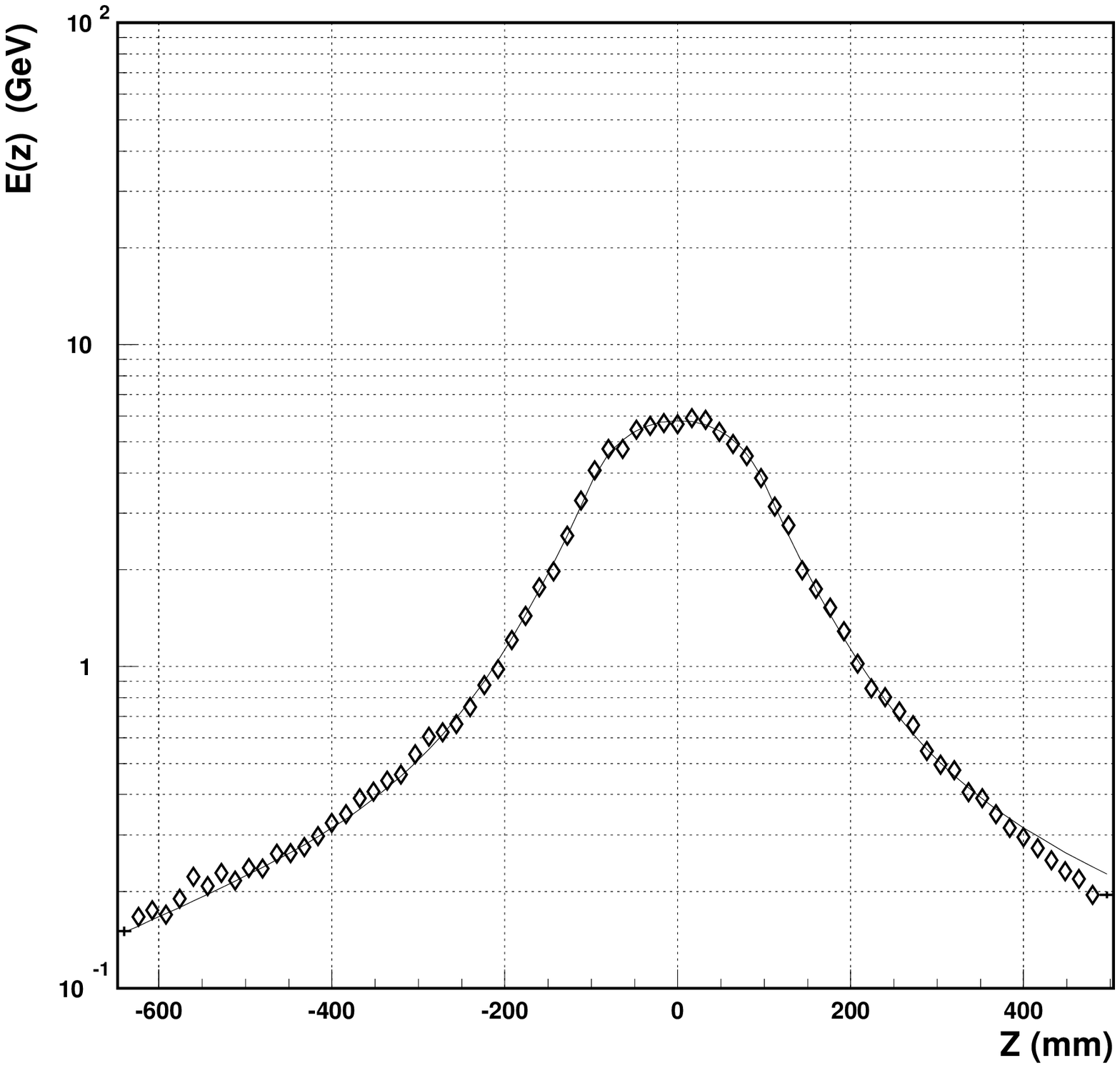,width=0.4\textwidth,height=0.2\textheight}}
        \\
\mbox{\epsfig{figure=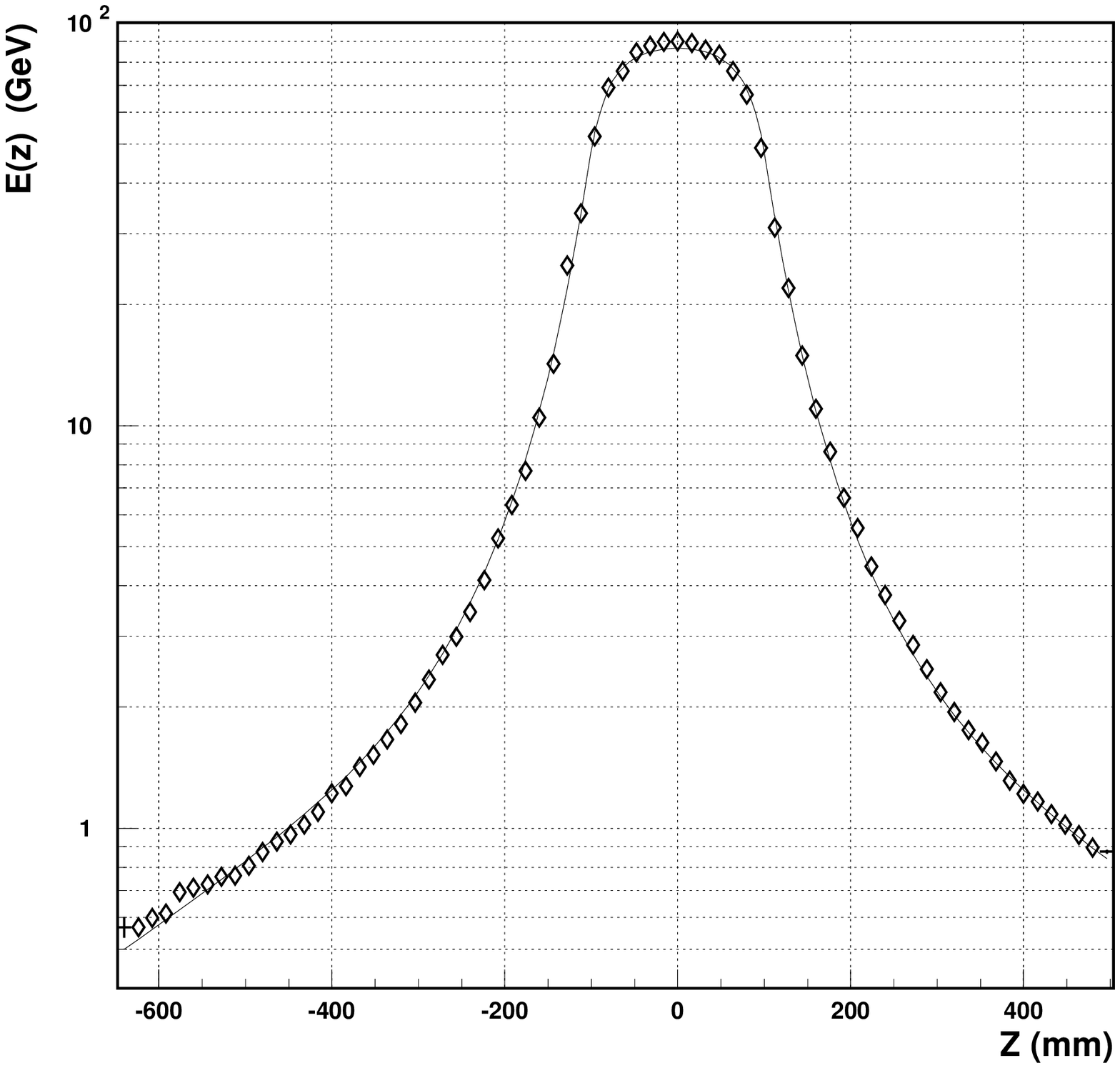,width=0.4\textwidth,height=0.2\textheight}}
& 
\\
        \end{tabular}
     \end{center}
      \caption{
        Energy depositions of 100 GeV pions in towers 
        of depth segments 1 -- 4  
        as a function of the $z$ coordinate:
        top left is for depth segment 1,
        top right is for depth segment 2,
        middle left is for depth segment 3,
        middle right is for depth segment 4, 
        bottom left is for over all calorimeter.
        Only statistical errors are shown.
       \label{fig:f5-001}}
\end{figure*}

Figure \ref{fig:f5-001} shows the energy depositions in towers  
for depth segments 1 -- 4 as a function of the $z$ coordinate of the 
center of the tower and for entire calorimeter.
Here the coordinate system is linked to the incident particle direction
where $z=0$ is the coordinate of the beam impact points at the 
calorimeter front face.
With fine parallel displacements of the beam between $- 360$ mm 
and $200$ mm we expand the tower coordinate range 
from $- 760$ mm to $600$ mm.
To avoid edge effects, we present tower energy depositions in the range 
from $- 650$ mm to $500$ mm.
The tower energy depositions shown in 
Figures \ref{fig:f5-001}
span a range of about three orders of magnitude.
The plateau for $|z| < 100$ mm ($h/2$) and the fall-off at large $|z|$
are apparent.
We used the distributions in Figs.\ \ref{fig:f5-001}
to extract the underlying 
mar\-gi\-nal densities function for four depth segments of the calorimeter
and for the entire calorimeter.
The solid curves in these figures
are the results of the fit with equations (4) and  (\ref{e023}).
The fits typically differ from the experimental distribution by less
than $5\%$.

\begin{table*}[tbh]
\caption{
        The parameters $a_i$ and $\lambda_i$ obtained by fitting 
        the transverse shower profiles for four depth segments and 
        the entire 
        calorimeter at 100 GeV.
         \label{Tb2}}
\begin{center}
\begin{tabular}{@{}|@{}c@{}|@{~}c@{~}|@{~}c@{~}|@{~}c@{~}|@{~}c@{~}|@{~}c@{~}|@{~}c@{~}|@{}}
\hline  
        $x$($\lambda_{\pi}^{Fe}$) 
        & $a_1$ 
        & $\lambda_1$(mm) 
        & $a_2$ 
        & $\lambda_2$(mm) 
        & $a_3$ 
        & $\lambda_3$(mm) 
\\
\hline
\hline
0.6
&$0.88\pm0.07$ 
&$17\pm2$ 
&$0.12\pm0.07$ 
&$48\pm14$ 
&$0.004\pm0.002$ 
&$430\pm240$
\\
\hline
2.0
&$0.79\pm0.06$ 
&$25\pm2$ 
&$0.20\pm0.06$ 
&$52\pm6$ 
&$0.014\pm0.006$ 
&$220\pm40$ 
\\
\hline
3.8
&$0.69\pm0.03$ 
&$32\pm8$ 
&$0.28\pm0.03$ 
&$71\pm13$ 
&$0.029\pm0.005$ 
&$280\pm30$ 
\\
\hline
6.0
&$0.41\pm0.05$ 
&$51\pm10$ 
&$0.52\pm0.06$ 
&$73\pm18$
&$0.07\pm0.03$ 
&$380\pm140$ 
\\ 
\hline
all
&$0.78\pm0.08$ 
&$23\pm1$ 
&$0.20\pm0.08$ 
&$58\pm4$
&$0.015\pm0.004$ 
&$290\pm40$ 
\\
\hline
\end{tabular}
\end{center}
\end{table*}

The parameters $a_i$ and $\lambda_i$, obtained by fitting, are listed in 
Table~\ref{Tb2}. 
The $a_i$ and $\lambda_i$ parameters demonstrate 
linear behaviour as a function of $x$:
$a_i (x) = \alpha_i + \beta_i  x$ and 
$\lambda_i (x) = \gamma_i + \delta_i  x$.
The values of the parameters 
$\alpha_i$, $\beta_i$, $\gamma_i$ and $\delta_i$
are presented in Table \ref{Tb02}.

\begin{table*}[tbh]
\caption{
         The values of the parameters 
         $\alpha_i$, $\beta_i$, $\gamma_i$ and $\delta_i$.
         \label{Tb02}}
\begin{center}
\begin{tabular}{|c||c|c|||c||c|c|}
\hline
        
        & $\alpha_i$ 
        & $\beta_i$     ($1 / \lambda_{\pi}$)
        & 
        & $\gamma_i$    (mm) 
        & $\delta_i$    ($\mathrm{mm} / \lambda_{\pi}$) 
\\
\hline
\hline
        $a_1$
&       $0.99 \pm 0.06$ 
&       $- 0.088 \pm 0.015$ 
&       $\lambda_1$
&       $13 \pm 2$ 
&       $6 \pm 1$ 
\\
\hline
        $a_2$
&       $0.04 \pm 0.06$ 
&       $0.071 \pm 0.015$ 
&       $\lambda_2$
&       $42 \pm 10$ 
&       $6 \pm 4$ 
\\
\hline
        $a_3$
&       $- 0.001 \pm 0.002$ 
&       $0.008 \pm 0.002$ 
&       $\lambda_3$
&       $170 \pm 80$ 
&       $29 \pm 23$ 
\\
\hline
\end{tabular}
\end{center}
\end{table*}

\section{Radial Hadronic Shower Energy Density}

\begin{figure*}[tbph]
     \begin{center}
        \begin{tabular}{cc}
\mbox{\epsfig{figure=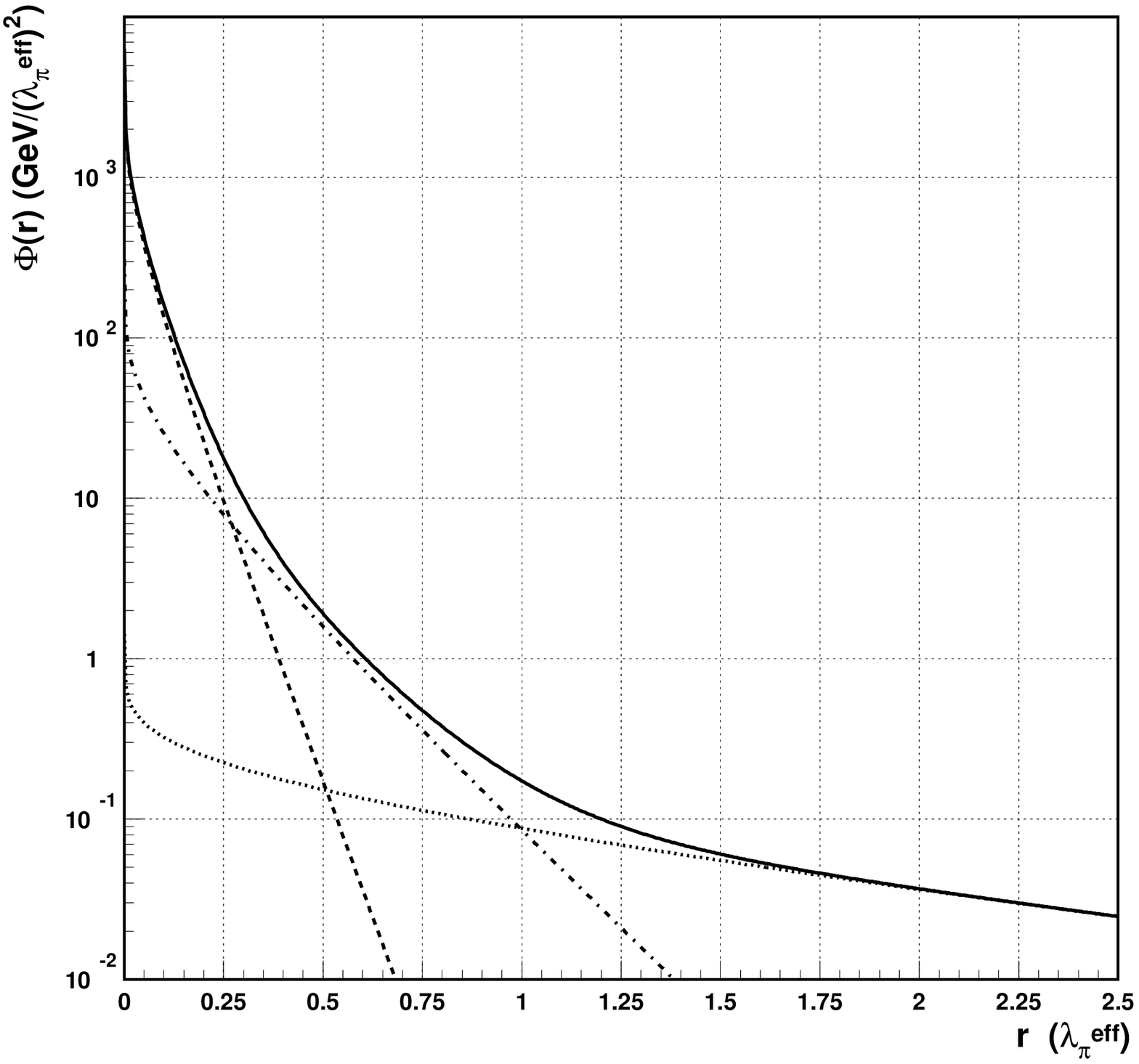,width=0.4\textwidth,height=0.2\textheight}}
&
\mbox{\epsfig{figure=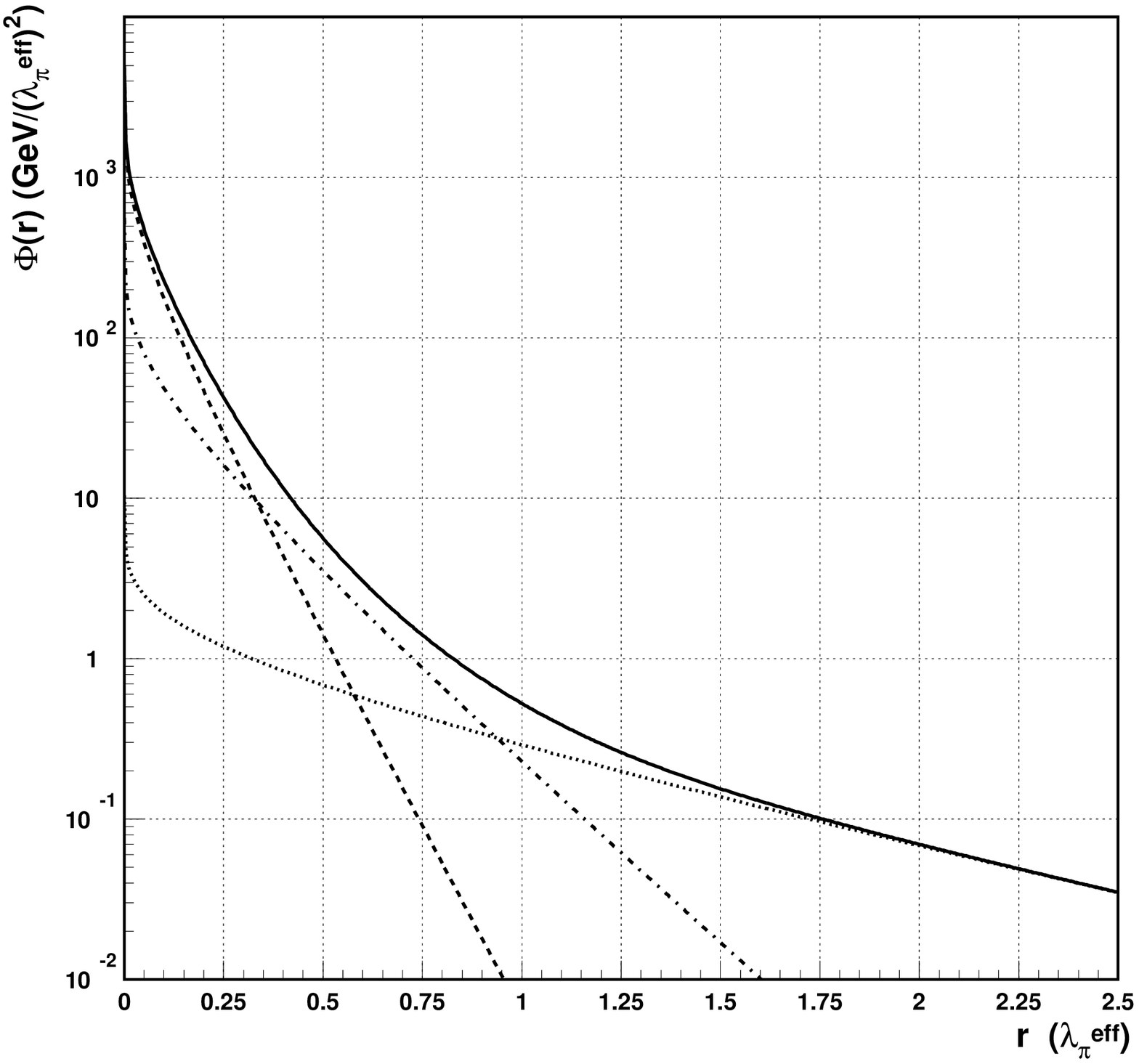,width=0.4\textwidth,height=0.2\textheight}}
        \\
\mbox{\epsfig{figure=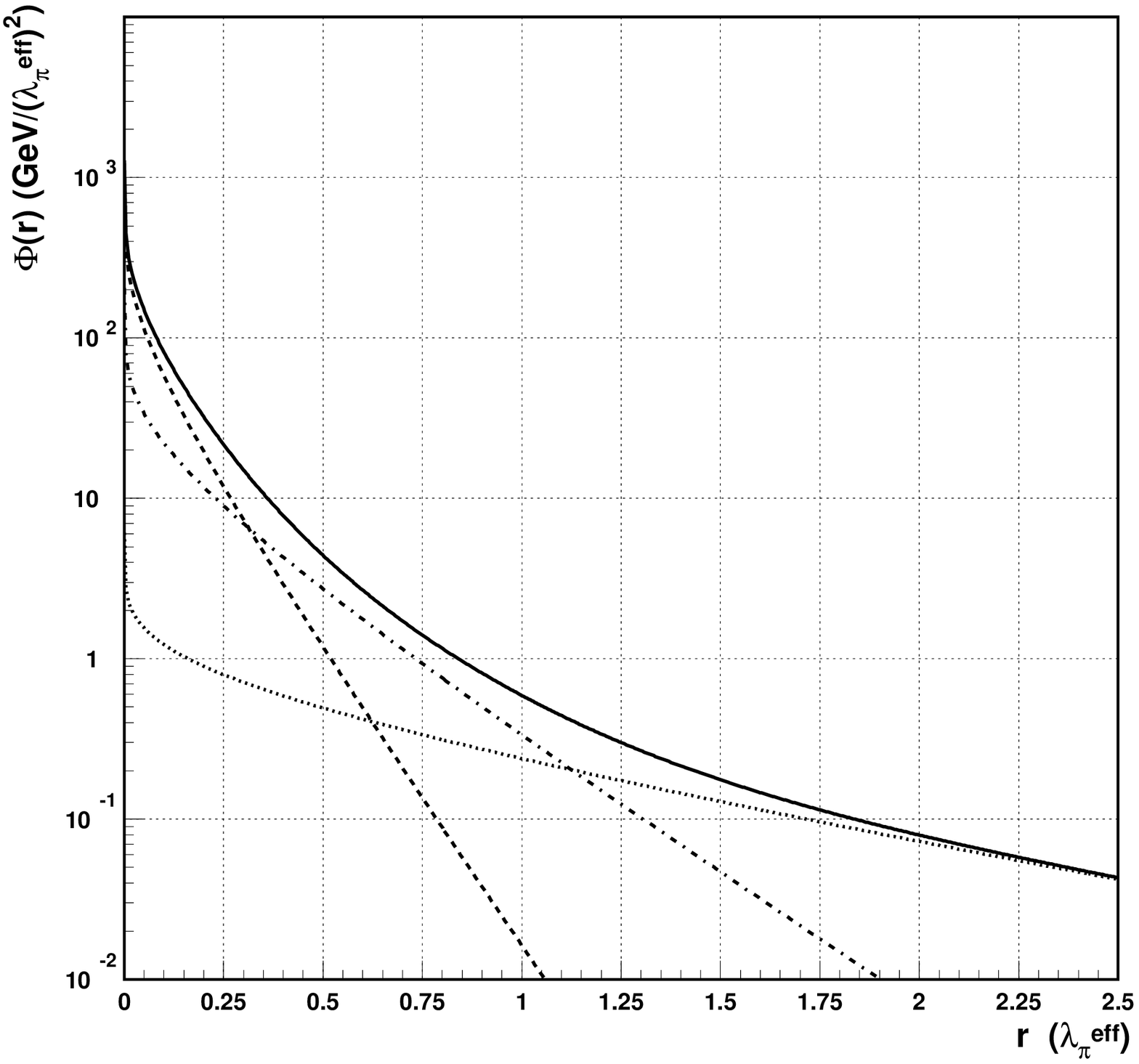,width=0.4\textwidth,height=0.2\textheight}}
&
\mbox{\epsfig{figure=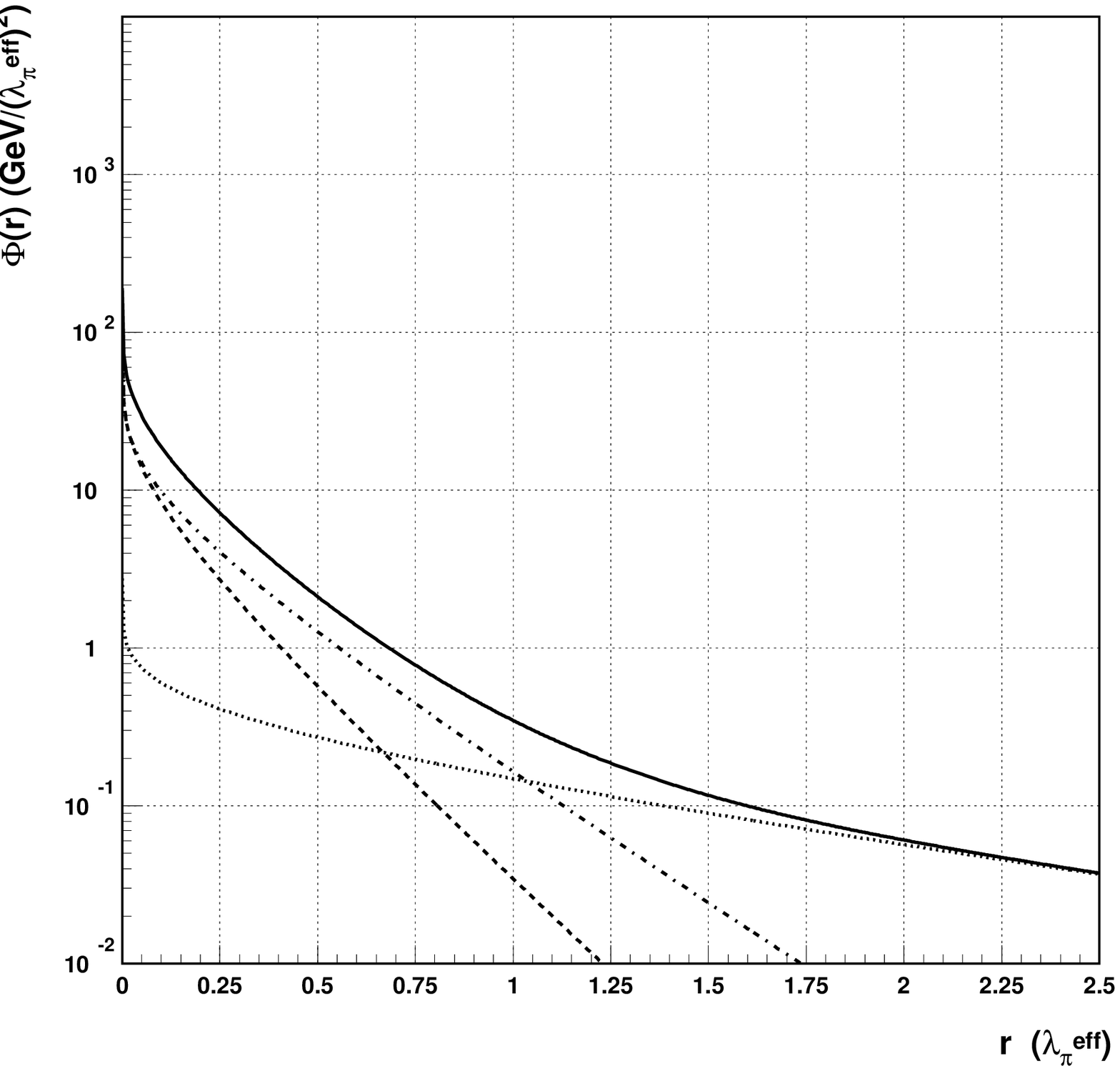,width=0.4\textwidth,height=0.2\textheight}}
        \\
\mbox{\epsfig{figure=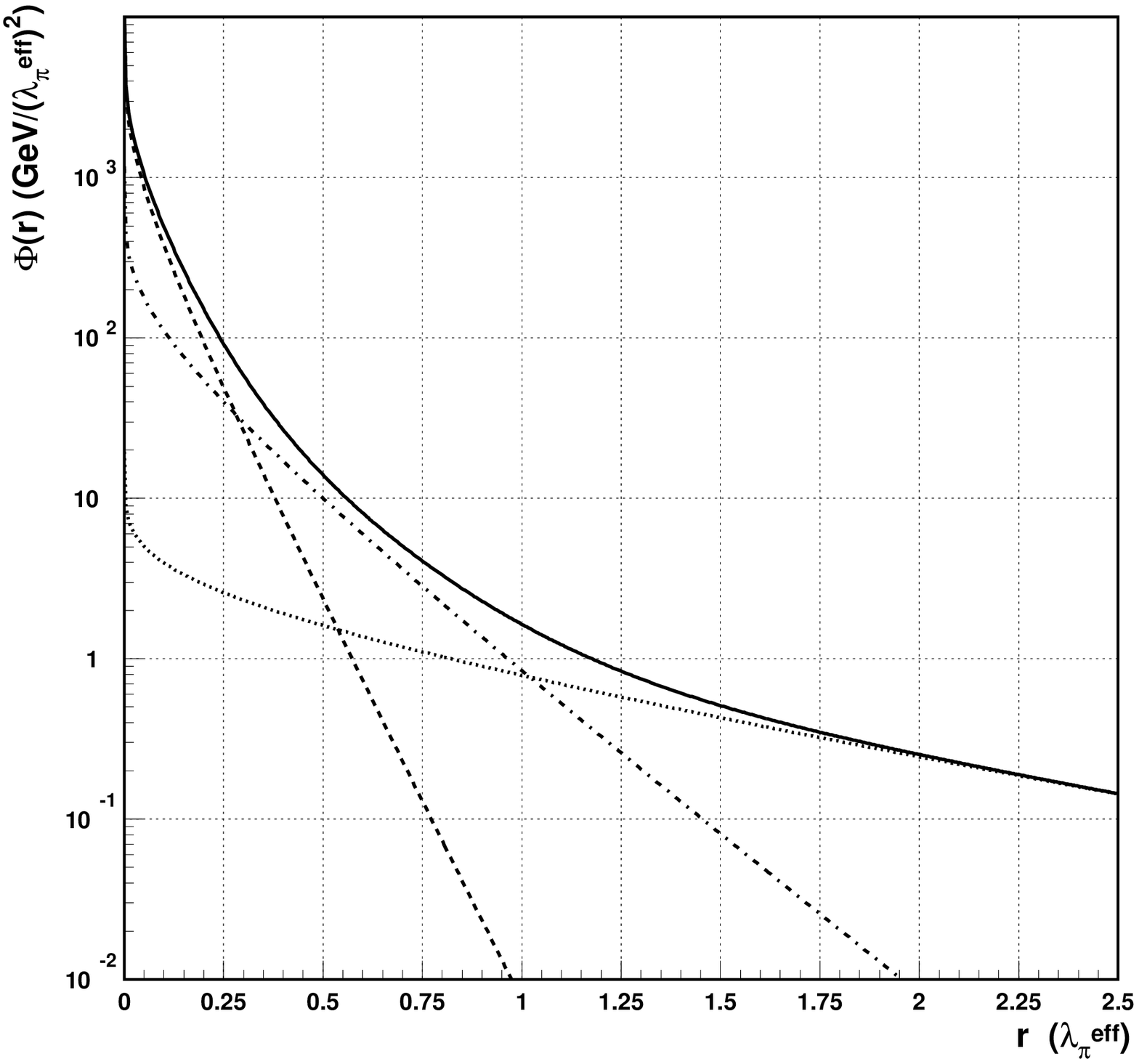,width=0.4\textwidth,height=0.2\textheight}}
&
\mbox{\epsfig{figure=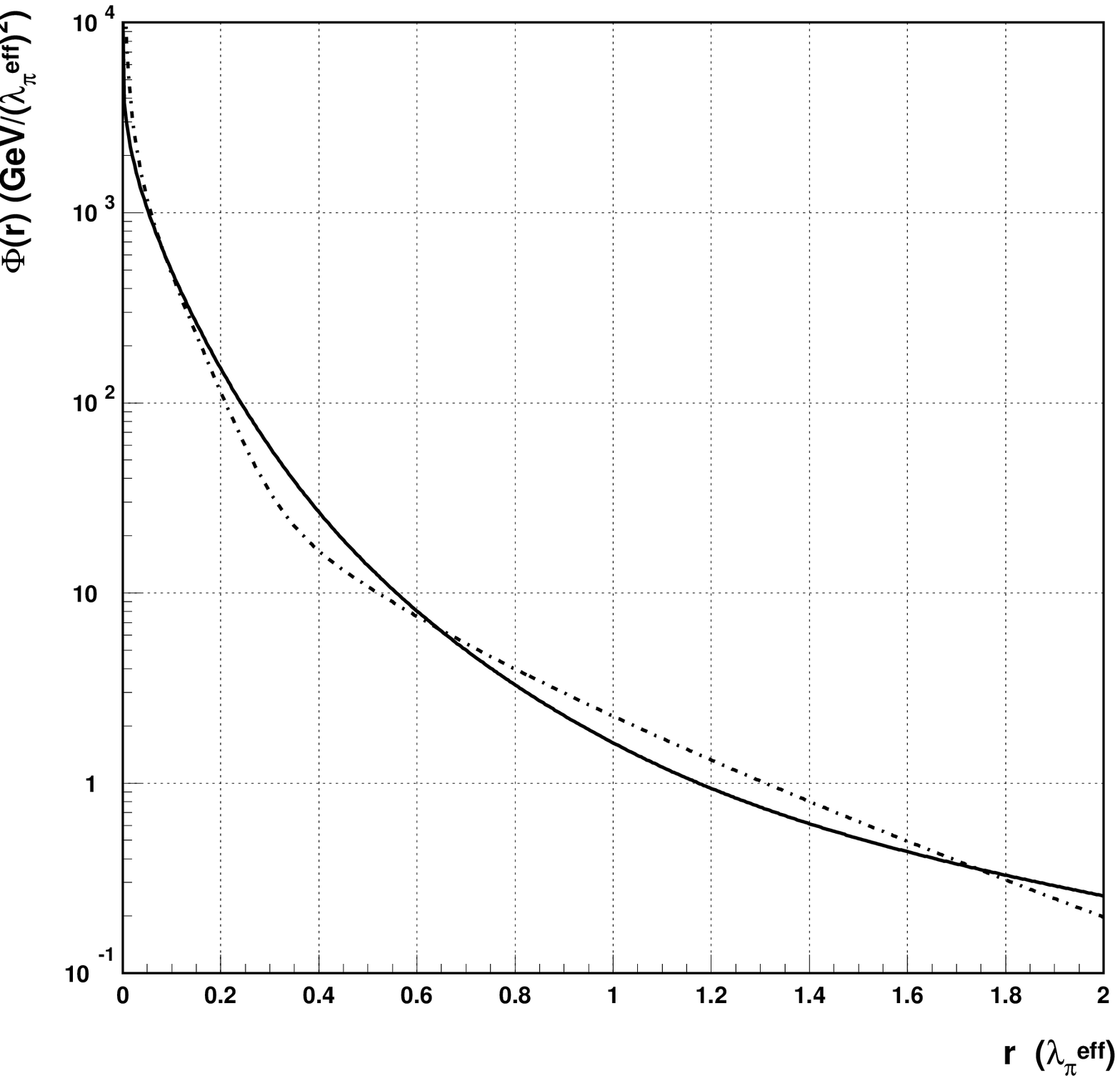,width=0.4\textwidth,height=0.2\textheight}}
        \\
        \end{tabular}
     \end{center}
      \caption{
        Radial energy density, $\Phi(r)$, as a function of $r$ for Tile 
        calorimeter
        for depth segments 1 -- 4:
        top left is for depth segment 1,
        top right is for depth segment 2,
        middle left is for depth segment 3,
        middle right is for depth segment 4, 
        bottom left is for entire calorimeter.
        The solid lines are the energy densities $\Phi(r)$,
        the dashed lines are the contribution from the first term from,
        the dash-dotted lines are the contribution from the second term,
        the dotted lines are the contribution from the third term.
        Bottom right: Comparison of the radial energy densities
        as a function of $r$ (in units of $\lambda_{\pi}^{eff}$) for
        Tile calorimeter (the solid line) and 
        lead-scintillating fiber calorimeter 
        (the dash-dotted line).
       \label{fig:f17-001}}
\end{figure*}

Using formula (\ref{e23}) and the values of the parameters 
$a_i$, $\lambda_i$, given in Table~\ref{Tb2}, 
we have determined the underlying radial 
hadronic shower energy density functions, $\Phi(r)$.
The results are shown in Figure \ref{fig:f17-001} for depth segments 1 -- 4 
and for the entire calorimeter.
The contributions of the three terms of $\Phi(r)$ are also shown.

The function $\Phi(r)$ for the entire calorimeter has been compared 
with the one for the lead-scintillating fiber calorimeter 
of Ref.\ \cite{acosta92}, that has
about the same effective nuclear interaction length for pions 
\cite{budagov97}.
The two radial density functions are rather similar as seen 
in Fig.\ \ref{fig:f17-001} (bottom right).
The lead-scintillating fiber calorimeter density function 
$\Phi(r)$ was obtained from 
a 80 GeV $\pi^{-}$ grid scan at an angle 
of $2^{\circ}$ with respect to the fiber direction.
For the sake of comparing the radial density functions of the 
two calorimeters,
the distribution from \cite{acosta92} was normalised to the $\Phi(r)$ of the
Tile calorimeter.
Precise agreement between these functions
should not be expected because of 
the effect of the different absorber materials used in the two detectors,
the values of $e/h$ are different, as is hadronic activity of showers because
fewer neutrons are produced 
in iron than in lead. 

\section{Radial Containment}

The parametrization of the radial density function, $\Phi(r)$,
was integrated to yield the shower containment as a function 
of the radius
$I(r) = E_{0} - \frac{E_{0} r}{B}\ \sum_{i=1}^{3}\ a_{i}\ 
K_1 (r/{\lambda}_{i})$,
where $K_{1}$ is the modified Bessel function.
The approximation of the data for the radii of cylinders for given shower 
containment ($90\%$, $95\%$ and $99\%$) as a function of depth 
with linear fits are
$r(90\%) = (85 \pm 6) + (37 \pm 3)x$,
$r(95\%) = (134 \pm 9) + (45 \pm 3)x$, 
$r(99\%) = (349 \pm 7) + (22 \pm 2)x$ (mm).
The centers of depth segments, 
$x$, are given in units of $\lambda_{\pi}^{Fe}$.

\section{Longitudinal Profile}

\begin{figure*}[tbph]
     \begin{center}
        \begin{tabular}{c}
\mbox{\epsfig{figure=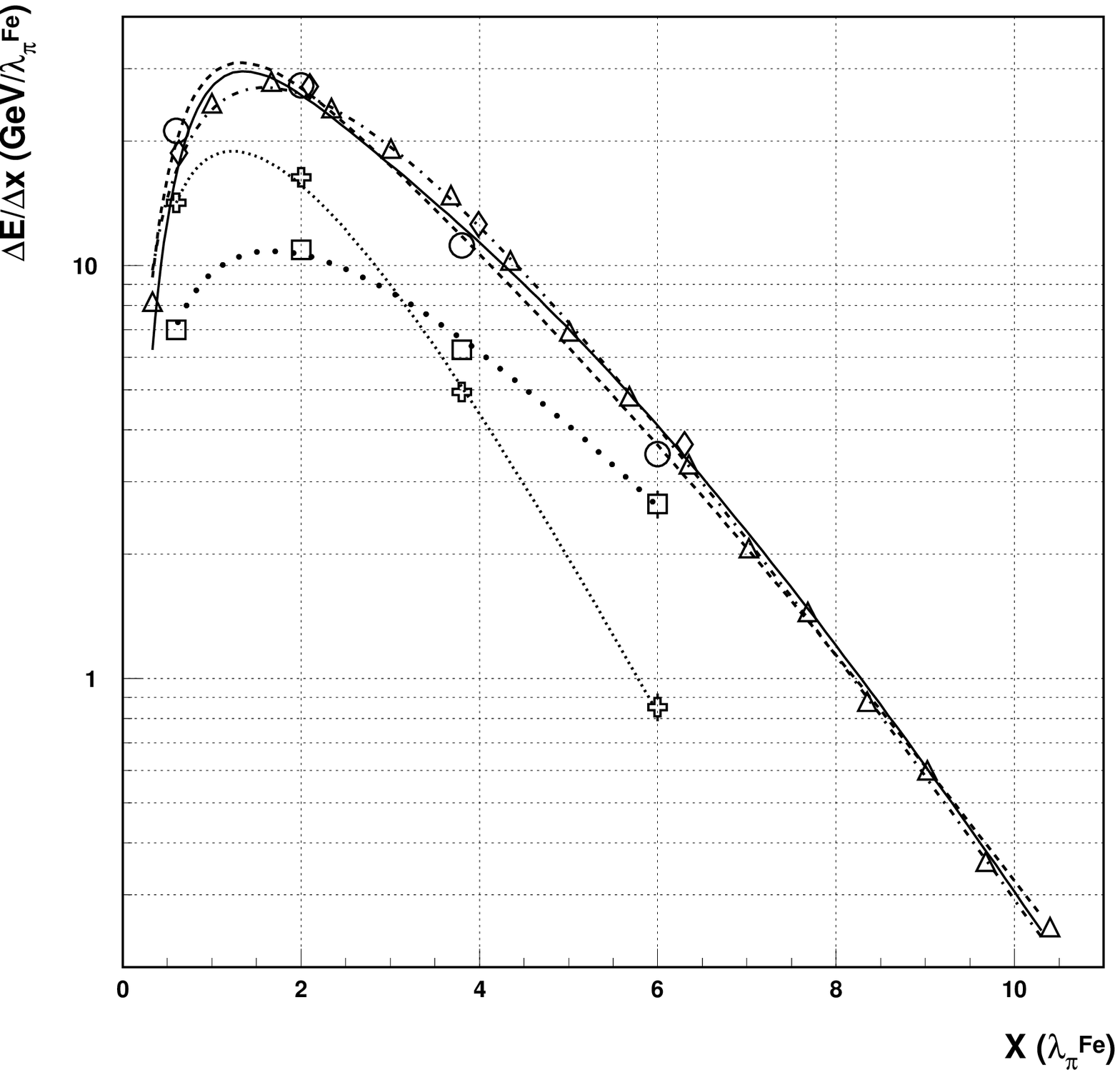,width=0.8\textwidth,height=0.2\textheight}} 
        \\
\mbox{\epsfig{figure=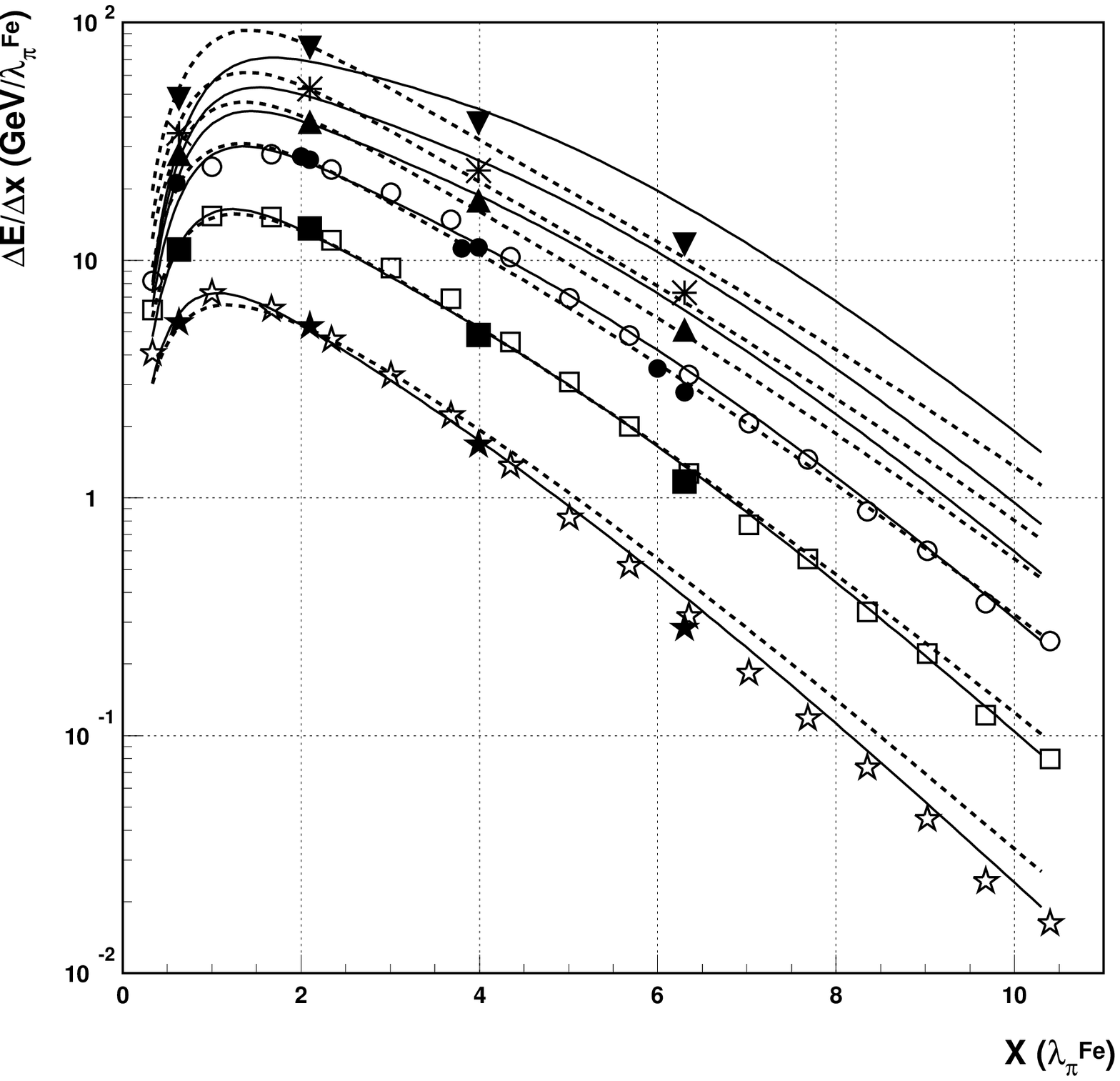,width=0.8\textwidth,height=0.2\textheight}} 
        \\
        \end{tabular}
     \end{center}
       \caption{
        Top: 
	The longitudinal profile (circles)
        of the hadronic shower at 100 GeV as a function of the
        longitudinal coordinate $x$ in units of $\lambda_\pi^{Fe}$.
        Open triangles are data from the calorimeter of Ref.$^7$ 
        diamonds are the Monte Carlo (GEANT-FLUKA) predictions. 
        The dash-dotted line is the fit by function 
	(6),
        the solid line is calculated with function 
	(7) 
        with parameters from Ref.$^7$ 
        the dashed line is calculated with function 
	(7) 
        with parameters from Ref.$^8$ 
        The electromagnetic and hadronic components of the shower
        (crosses and squares), together with their fits using
        (6),
	are discussed in Section 8.
        Bottom: 
        Longitudinal profiles of the hadronic showers 
	from 20 GeV (open stars),
        50 GeV (open squares) and 100 GeV (open circles) 
        pions as a function of the longitudinal coordinate $x$ in units 
        of $\lambda_I$ for a conventional iron-scintillator 
        calorimeter $^7$
	and of 20 GeV (black stars),
        50 GeV (black squares), 100 GeV (black circles), 
        150 GeV (black up triangles),
        200 GeV (asterisks), 300 GeV (black down triangles)
        for pions at $20^\circ$  
        and of 100 GeV (black circles) for pions at $10^\circ$
        for Tile iron-scintillator calorimeter. 
        The solid lines are calculated with function 
	(7)
        with parameters from Ref.$^7$
	and the dashed lines
        are with parameters from Ref.$^8$
       \label{fig:f15}}
\end{figure*}

Our values of  $\Delta E / \Delta x$
together with the data of \cite{hughes90}
and Monte Carlo predictions (GEANT-FLUKA + MICAP) \cite{juste95} are
shown
in Fig.\ \ref{fig:f15} (top).
The longitudinal energy deposition
for our calorimeter is in good agreement with that of a 
conventional iron-scintillator calorimeter.
The longitudinal profile may be approximated
using two parametrizations.
The first form is 
\begin{equation}
        \frac{dE(x)}{dx} = 
                \frac{E_{f}\ \beta^{\alpha + 1}}{\Gamma (\alpha + 1)}\
                x^{\alpha}\ e^{-\beta x}
\label{elong00}
\end{equation}
where $E_{f} = E_{beam}$, and $\alpha$ and $\beta$ are free parameters.
Our data at 100 GeV and those of Ref.\ \cite{hughes90} at 100 GeV were
jointly fit to this expression; the fit is shown in 
Fig.\ \ref{fig:f15} (top).
The second form is the analytical representation of the
longitudinal shower profile from the front of the calorimeter
\begin{eqnarray}
        \frac{dE (x)}{dx} & = & 
                N\ 
                \Biggl\{
                \frac{w\ X_{0}}{a}\ 
                \biggl( \frac{x}{X_{0}} \biggr)^a\ 
                e^{- b \frac{x}{X_{0}}}\
                {}_1F_1 \biggl( 1,\ a+1,\ 
                \biggl(b - \frac{X_{0}}{\lambda_I} \biggr)\ \frac{x}{X_{0}}
                \biggr) 
                \nonumber \\
                & & + \ 
                \frac{(1 - w)\ \lambda_I}{a}\ 
                \biggl( \frac{x}{\lambda_I} \biggr)^a\ 
                e^{- d \frac{x}{\lambda_I}}\
                {}_1F_1 \biggl( 1,\ a+1,\
                \bigl( d - 1 \bigr)\ \frac{x}{\lambda_I} \biggr)
                \Biggr\} , 
\label{elong2}
\end{eqnarray}
where $a,\ b,\ d,\ w$ are parameters,
${}_1F_1$ is the confluent hypergeometric function. 
Here the depth variable, $x$, is the depth in
equivalent $Fe$, $X_0$ is the radiation length in $Fe$
and in this case $\lambda_I$ is $\lambda_\pi^{Fe}$.
The normalisation factor $N  = E_{beam}/\lambda_I\ \Gamma (a)\
(w\ X_{0}\ b^{-a} + (1-w)\ \lambda_I\ d^{-a})$. 
This form was suggested in \cite{kulchitsky98} and derived by integration  
over the shower vertex positions of the longitudinal shower development
from the shower origin.
For the para\-me\-trization of longitudinal shower development 
from the shower origin, the well known parametrization 
suggested by Bock et al.\ \cite{bock81} has been used.
We compare the form (\ref{elong2}) 
to the experimental points at 100 GeV using the
parameters calculated in Refs.\ \cite{bock81} and \cite{hughes90}. 
Note
that now we are not performing a fit but checking how well the general
form (\ref{elong2}) 
together with two sets of parameters for iron-scintillator
calorimeters describe our data. 
As shown in Fig.\ \ref{fig:f15} (top), both sets of
parameters work rather well in describing the 100 GeV data.
Turning next to the longitudinal shower development at different
energies, in Fig.\ \ref{fig:f15} (bottom) our values of  
$\Delta E / \Delta x$ 
for 20 -- 300 GeV together with the data from \cite{hughes90} are shown.
The solid and dashed lines are calculations with function (\ref{elong2})
using parameters from \cite{hughes90} and \cite{bock81}, respectively.
Again, we observe reasonable agreement between our data and the 
corresponding data for conventional iron-scintillator 
calorimeter on one hand,  and between data and the 
parametrizations described above.
Note that the fit in \cite{hughes90} has been performed in the energy 
range from 10 to 140 GeV; hence 
the curves for 200 and 300 GeV should be considered as extrapolations.
It is not too surprising that at these energies the agreement is
significantly worse, particularly at 300 GeV. 
In contrast, the parameters of \cite{bock81}
were derived from data spanning the range 15 -- 400 GeV, and are in much
closer agreement with our data.

\section{The parametrization of Hadronic Showers}

The three-dimensional parametrization for spatial hadronic shower
 development is
        $$\Psi (x, r) = 
                \frac{dE(x)}{dx} \cdot 
                \frac{ 
                \sum_{i=1}^{3} \ \frac{a_{i} (x)}{\lambda_{i} (x)}\
                K_{0} \bigl( \frac{r}{\lambda_{i} (x)} \bigr)
                }{
                2\pi\ \sum_{i=1}^{3} a_{i}(x) \lambda_{i} (x)}\ ,$$ 
where $dE  (x)/ dx$, defined by equation (\ref{elong2}),
is the longitudinal energy deposition.

\section{Electromagnetic Fraction of Hadronic Showers}

\begin{figure*}[tbph]
     \begin{center}
        \begin{tabular}{c}
\mbox{\epsfig{figure=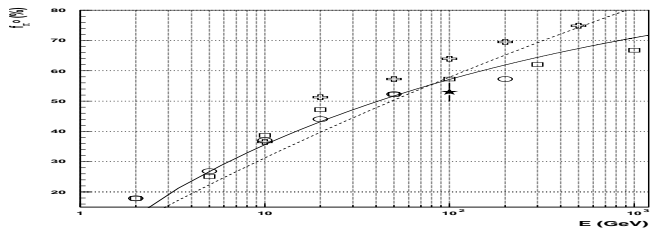,width=0.8\textwidth,height=0.2\textheight}} 
        \\
        \end{tabular}
     \end{center}
       \caption{
        The fraction $f_{\pi^{0}}$ in 
        hadronic showers versus the beam energy.
        The star is our data, 
        the solid curve is the Groom parametrization, 
        the dashed curve is the Wigmans parametrization,
        squares are the GEANT-CALOR predictions,
        circles are the GEANT-GHEISHA predictions and
        crosses are the CALOR predictions.
       \label{fig:f150-1}}
\end{figure*}

Following \cite{acosta92},
we assume that the electromagnetic part of a hadronic shower is the
prominent central core, which in our case is the first term in the
expression (\ref{e23}) for the radial energy density function, $\Phi(r)$.
Integrating $f_{\pi^0}$ over $r$ we get
$f_{\pi^{0}} = a_1 \lambda_1 / \sum_{i=1}^{3} a_i \lambda_i$.
For the entire Tile calorimeter this value is $(53 \pm 3) \%$
at 100 GeV.
The observed $\pi^{0}$ fraction, $f_{\pi^{0}}$, is related to the 
intrinsic actual fraction, $f^{\prime}_{\pi^{0}}$, by the equation
$f_{\pi^{0}}(E) =
        \frac{e\ E^{\prime}_{em}}{e\ E^{\prime}_{em} + h\ E^{\prime}_{h}} = 
                \frac{ e/h \cdot f^{\prime}_{\pi^{0}}(E)}{(e/h - 1) 
                \cdot f^{\prime}_{\pi^{0}}(E) + 1}$,
where $E^{\prime}_{em}$ and $E^{\prime}_{h}$ are the intrinsic 
electromagnetic and hadronic parts of shower energy,
$e$ and $h$ are the coefficients of conversion of intrinsic electromagnetic 
and hadronic energies into observable signals, 
$f^{\prime}_{\pi^{0}} = E^{\prime}_{em}/(E^{\prime}_{em} + E^{\prime}_{h})$.
There are two analytic forms for the intrinsic $\pi^{0}$
fraction suggested by Groom \cite{groom90}
$f^{\prime}_{\pi^{0}}(E) = 1 - (E/E_{0}^\prime)^{m-1}$
and Wigmans \cite{wigmans88}
$f^{\prime}_{\pi^{0}}(E) =  k \cdot ln{(E/E_0^\prime)}$,
where $E_{0}^{\prime} = 1$ GeV, $m = 0.85$ and $k = 0.11$.
We calculated $f_{\pi^{0}}$ using the value $e/h = 1.34\pm0.03$ 
for our calorimeter 
\cite{budagov95-72}
and obtained the curves shown in Fig.\ \ref{fig:f150-1}.
Our result at 100 GeV is compared 
to the modified Groom and Wigmans parametrizations and to results from
the Monte Carlo codes 
CALOR \cite{gabriel94}, 
GEANT-GEISHA and GEANT-CALOR 
(the latter code is an implementation of
CALOR89 differing from GEANT-FLUKA only for 
hadronic interactions below 10 GeV). 
As can be seen from Fig.\ \ref{fig:f150-1}, our calculated 
value of $f_{\pi^{0}}$ is about one standard deviation lower than 
two of the Monte Carlo results 
and the Groom and Wigmans parametrizations.
The fractions $f_{\pi^{0}} (r)$ for the entire calorimeter 
and for depth segments 1 -- 3 amount to about $90\%$ as $r \rightarrow 0$ 
and decrease to about $1\%$ as $r \rightarrow \lambda_{\pi}^{eff}$.
However for depth segment 4 the value of $f_{\pi^{0}} (r)$ amounts 
to only $50\%$ as $r \rightarrow 0$ and decreases slowly to about $10\%$ as
$r \rightarrow \lambda_{\pi}^{eff}$. 
The values of $f_{\pi^{0}}(x)$  as a function of $x$ are fitted by 
$f_{\pi^{0}}(x) = (75\pm2) - (8.4\pm0.4) x$ ($\%$).
Using the values of $f_{\pi^{0}}(x)$ and energy depositions for various 
depth segments, we obtained the contributions from the electromagnetic and 
hadronic parts of hadronic showers in Fig.\ \ref{fig:f15} (top).
The curves represent a fit to the electromagnetic
and hadronic components of the shower using equation (\ref{elong00}).
$E_{f}$ is set equal to $f_{\pi^{0}} E_{beam}$
for the electromagnetic fraction and
$(1 - f_{\pi^{0}}) E_{beam}$ for the hadronic fraction.
The electromagnetic component of a hadronic shower rise and decrease 
more rapidly than the hadronic one 
($\alpha_{em} = 1.4\pm0.1$, $\alpha_{h} = 1.1\pm0.1$, 
$\beta_{em} = 1.12\pm0.04$, $\beta_{h} = 0.65\pm0.05$).
The shower maximum position
($x_{max} = (\alpha / \beta )\ \lambda_{\pi}^{Fe}$)
occurs at a shorter distance from the calorimeter front face 
($x_{max}^{em} = 1.23\ \lambda_{\pi}^{Fe}$, 
 $x_{max}^{h} = 1.85\ \lambda_{\pi}^{Fe}$).
At depth segments greater than $4\ \lambda_{\pi}^{Fe}$, the hadronic
fraction of the shower begins to dominate.
This is natural since 
the energy of the secondary hadrons is too low to permit significant 
pion production.

\section{Summary and Conclusions}

We have investigated the lateral development of hadronic showers
using 100 GeV pion beam data at an incidence angle of
$\Theta = 10^{\circ}$ for impact points $z$ in the range from 
$- 360$ to $200$ mm
and the longitudinal development of hadronic showers 
using 20 -- 300 GeV pion beams at an incidence angle of 
$\Theta = 20^{\circ}$.
We have obtained for four depth segments and for the entire calorimeter:
energy depositions in towers;
underlying radial energy densities;
the radii of cylinders for a given shower containment fraction;
the fractions of the electromagnetic and hadronic parts of a shower;
differential longitudinal energy deposition.
The three-dimensional parametrization of hadronic showers 
that we obtained allows direct use in any application that requires
volume integration of shower energy depositions  
and position reconstruction.

\section{Acknowledgements}

This paper is the result of the efforts of many people from the ATLAS
Collaboration.
The authors are greatly indebted to the entire Collaboration
for their test beam setup and data taking.
We are grateful to the staff of the SPS, and in
particular to Konrad Elsener, for the excellent beam conditions and
assistance provided during our tests.


\end{document}